\documentclass{article}

\usepackage[utf8]{inputenc}

\usepackage{bbold}
\usepackage{microtype}
\usepackage{graphicx}
\usepackage{booktabs} 
\usepackage{amsthm}
\usepackage{amsfonts}
\usepackage{amsmath}
\usepackage{amssymb}
\usepackage{scrextend}
\usepackage{hyperref}

\usepackage{subcaption}
\usepackage{graphicx}
\usepackage[english]{babel}
\usepackage{pgfplots}
\usepackage{multirow}

\usepackage{mathtools}
\usepackage{url}
\usepackage{comment}
\usepackage[numbers]{natbib}

\theoremstyle{plain}
\newtheorem{theorem}{Theorem}

\newtheorem{corollary}{Corollary}[section]
\newtheorem{definition}{Definition}[section]
\pgfplotsset{compat=1.14}
\newcommand{\norm}[1]{\left\lVert#1\right\rVert}

\usepackage{amsmath}
\usepackage{amsfonts}
\usepackage{nicefrac}

\usepackage[accepted]{icml2021_style/icml2021}

\usepackage{todonotes}
\usepackage{fixme}
\fxsetup{status=draft}
\FXRegisterAuthor{tv}{atv}{TV}

\begin{document}
	
	\twocolumn[
	\icmltitle{BRR: Preserving Privacy of Text Data Efficiently on Device}
	
	\icmlsetsymbol{equal}{*}
	
	\begin{icmlauthorlist}
		\icmlauthor{Ricardo Silva Carvalho}{sfu}
		\icmlauthor{Theodore Vasiloudis}{amz}
		\icmlauthor{Oluwaseyi Feyisetan}{amz}
	\end{icmlauthorlist}
	
	\icmlaffiliation{sfu}{Simon Fraser University}
	\icmlaffiliation{amz}{Amazon.com}
	
	\icmlcorrespondingauthor{Ricardo Silva Carvalho}{rsilvaca@sfu.ca}
	\icmlcorrespondingauthor{Theodore Vasiloudis}{thvasilo@amazon.com}
	
	\icmlkeywords{Differential Privacy, NLP, Efficiency}
	
	\vskip 0.3in
	]
	
	\printAffiliationsAndNotice{}
	
	\begin{abstract}
		With the use of personal devices connected to the Internet for tasks such as searches and shopping becoming ubiquitous, ensuring the privacy of the users of such services has become a requirement in order to build and maintain customer trust. While text privatization methods exist, they require the existence of a trusted party that collects user data before applying a privatization method to preserve users' privacy.
		In this work we propose an efficient mechanism to provide metric differential privacy for text data on-device. With our solution, sensitive data never leaves the device and service providers only have access to privatized data to train models on and analyze.
		We compare our algorithm to the state-of-the-art for text privatization, showing similar or better utility for the same privacy guarantees, while reducing the storage costs by orders of magnitude, enabling on-device text privatization.
	\end{abstract}
	
	\section{Introduction}

As users interact with more and more edge devices through text and speech, organizations are able
to train models on the data received from users to offer better services and drive business. However,
research has demonstrated that machine learning models trained on sensitive data can be attacked,
and sensitive information about individuals extracted from the original training data\cite{shokri2017membership}.

In order to build and maintain customer trust, companies must protect the privacy of their users, so
that they will continue to use and be delighted by the products offered, without mistrust towards
the platform and the data holder. The best way to ensure that is by sending no sensitive data
to the service provider at all, known as the \textit{zero-trust} model\cite{dwork2006calibrating}. That way sensitive data never
leaves the device, and users can be assured that even in the event of a data breach of the
service provider, their private data will not be compromised.

Differential Privacy (DP)~\cite{dwork2006calibrating} has emerged as the most well-established technique to provide privacy guarantees
for individuals. However, DP was originally designed to deal with continuous data, while categorical and text data pose
an additional challenge~\cite{dwork2014algorithmic}. DP works by applying noise to inputs, and for a naive approach to work for text in the \textit{zero-trust} model, it would have to ensure that
a word can be replaced by \emph{any} other word in the vocabulary, potentially ruining the utility of the algorithm.

To deal with this restriction, previous research has proposed applying a privatization step in the continuous
embedding space instead~\cite{natasha2018obfuscation}, using a generalization of DP for metric spaces called \emph{metric-DP} \cite{chatzikokolakis2013metricdp}. These methods work by retrieving the embedding vector of the word we want to privatize, adding noise,
then replacing the word with the one closest to the new noisy vector in the embedding space.
However, these methods assume a central authority that gathers the data of all the users and applies the
privatization mechanism to generate a new privatized dataset, before training downstream NLP models. 
If we try using these mechanisms on-device, the space cost would be prohibitive. The embedding vectors plus the nearest neighbor index necessary
to retrieve the perturbed words from the noisy embedding can reach several gigabytes.

In this paper we propose an approach that allows for zero-trust text data privatization, ensuring that sensitive
data never leave the device of the user, improving trust to the service.
The key component of our approach is to use binary embedding vectors~\cite{tissier2019near} to dramatically shrink the space
and computational costs of storing and querying the word embeddings, while maintaining semantic meaning. Using binary word embedding vector representations, we  propose a text privatization
mechanism based on the randomized response~\cite{wang2016rr}, and prove that the mechanism satisfies metric-DP. 

In summary, our contributions are the following:

\begin{itemize}
	\item We propose a zero-trust algorithm for on-device text privatization, using binary embeddings and randomized response.
	\item A proof that our mechanism satisfies metric-DP is included, more specifically for Hamming distance.
	\item We develop theoretical methods for comparing metric-DP mechanisms that use different metrics, allowing consistent privacy-utility evaluation.
	\item Finally, our empirical evaluation demonstrates the computational advantages of the
		approach compared to the state-of-the-art, while maintaining better or similar utility.
\end{itemize}

	\section{Related Work}\label{sec:rel_work}

\citet{natasha2018obfuscation} were one of the first to use metric differential privacy on text data. They focused on the ``bag of words'' representation of documents and applied the Earth Mover's metric to obtain privatized bags, being also the first work to perform individual word privatization in the context of metric differential privacy. Following this context, the Madlib\footnote{We refer to this algorithm with the name used in \cite{madlibBlog}} mechanism \cite{madlib2020} adds noise to embedding vectors of words, working on the Euclidean space and adding Laplacian noise. After introducing noise, the mechanism outputs the word that is closest to the noisy vector in the embedding space. The algorithm presented in \cite{poincareMDP2019} is a follow-up to \cite{madlib2020} although it appeared later. This mechanism works in a hierarchical embedding space, where the embedding vector of an input word is perturbed with noise from a hyperbolic distribution. These works successfully illustrated the privacy-utility trade-off on metric differential privacy, and empirically showed that we can achieve reasonable privacy guarantees with the impact on the utility of downstream text models being dependent on the complexity of the downstream task. For example, the complex
question-answering task was more affected than binary classification.

\citet{poincareMDP2019} compare the hyperbolic mechanism to Madlib \cite{madlib2020}. However, since the two algorithms use different metric functions, the evaluation of privacy via only matching the $\varepsilon$ parameter of differential privacy can be improved. In this sense, \citet{poincareMDP2019} compares the privacy of the two mechanisms, looking at the probability of not changing a word after noise injection, i.e. the probability that the mechanism returns the exact same word used as input. Even though this notion can be intuitively seen as a level of indistinguishability, it cannot guarantee a fair comparison between mechanisms. In Section \ref{sec:comparingDPmec} we propose a method that ensures a more fair comparison, based on a privacy loss bound.

Finally, to the best of our knowledge, our work is the first to apply metric differential privacy on text data efficiently
on-device. This application scenario allows users to share only already privatized data, keeping their sensitive information local.
In this sense, our intended use is similar to the goals of Local Differential Privacy (LDP)\cite{dwork2006calibrating}. Nonetheless, previous work on LDP
focused on aggregate statistics, instead of individual word privatization. Examples are Google’s RAPPOR \cite{fanti2016building},
Apple’s DP distributed system \cite{appleDP} and Microsoft’s Private Collection of Telemetry Data \cite{ding2017collecting}. In the 
context of individual privatization of a given input, in our case word, using LDP would mean adding extremely amounts of noise, which
metric-DP relaxes by including distance metrics within the differential privacy guarantees.

	\section{Preliminaries}

Consider a user giving as input a word $w$ from a discrete fixed domain $\mathcal{W}$. For any pair of inputs $w$ and $w'$, we
assume a distance function $d: \mathcal{W} \times \mathcal{W} \rightarrow \mathbb{R}_{+}$, in a given space of representation of
these words. Specifically, we consider a word embedding model $\phi: \mathcal{W} \rightarrow \mathbb{R}^n$ will be used to
represent words, and the distance function can be a valid metric applicable to the embedding vectors.

Our goal is to select a word from $\mathcal{W}$, based on a given input, such that the privacy of the user, with respect to their
word choice, is preserved. From an attacker's perspective, the output of an algorithm working over an input $w$ or $w'$ will
become more similar as these inputs become closer according to the distance $d(w, w')$. In other words, if two words are close on the embedding space, they tend to generate similar results with same probabilities. 

With that in mind, we will work on Metric-Differential Privacy \cite{chatzikokolakis2013metricdp}, a privacy standard defined for
randomized algorithms with input from a domain $\mathcal{W}$ that are equipped with a distance metric $d: \mathcal{W} \times
\mathcal{W} \rightarrow \mathbb{R}_{+}$ satisfying the formal axioms of a metric. In this context, algorithms satisfying metric-DP
will have privacy guarantees that depend not only on the privacy parameter $\varepsilon$, but also on the particular distance
metric $d$ being used. 

\begin{definition} (Metric Differential privacy \cite{chatzikokolakis2013metricdp})\label{def:metric_dp} Given a distance metric
	$d: \mathcal{W} \times \mathcal{W} \rightarrow \mathbb{R}_{+}$, a randomized mechanism $\mathcal{M}: \mathcal{W} \rightarrow
	\mathcal{Y}$ is $\varepsilon d$-differentially private if for any $w, w' \in \mathcal{W}$ and all outputs $y \in \mathcal{Y}$ we
	have:
	\begin{gather}\label{eq:mDP}
	\Pr[\mathcal{M}(w) = y] \leq e^{\varepsilon d(w, w')} \Pr[\mathcal{M}(w') = y]
	\end{gather}
\end{definition}

Usually, on the standard definition of differential privacy\cite{dwork2006calibrating}, the privacy guarantees provided by
different mechanisms are compared by looking at the $\varepsilon$ value, such that mechanisms with same $\varepsilon$ give the
same privacy guarantee. For a fair evaluation on metric-DP using $\varepsilon$, we also have to consider the distance metrics used, since it also affects the privacy guarantees, as we can see from Definition~\ref{def:metric_dp}. We describe our theoretically motivated method to enable a fair privacy comparison of mechanisms with different metrics in Section \ref{sec:comparingDPmec}.

For Euclidean distance as metric, as discussed on Section~\ref{sec:rel_work}, the current state-of-the-art is the Madlib
mechanism. It uses the Laplace mechanism to add Laplacian noise to a given vector in order to obtain a private output.

\begin{algorithm}[tb]
	\caption{- \textbf{Madlib}: Word Privatization Mechanism for Metric Differential Privacy}
	\label{alg:madlib}
	\begin{algorithmic}[1]
		\INPUT Finite domain $\mathcal{W}$, input word $w \in \mathcal{W}$ and privacy parameter $\varepsilon$.\;\\
		\OUTPUT Privatized word $\hat{w}$.
		\STATE Compute embedding $\phi_w = \phi(w)$
		\STATE Perturb embedding to obtain $\hat{\phi}_w = \phi_w + N$ with noise density $p_N(z) \propto \exp(-\varepsilon \norm{z})$
		\STATE Obtain perturbed element: \\ $\hat{w} = argmin_{y \in \mathcal{W}} \norm{\phi(y) - \hat{\phi}_w}$
		\STATE Return $\hat{w}$
	\end{algorithmic}
\end{algorithm}

For a Euclidean metric $d: \mathcal{W} \times \mathcal{W} \rightarrow \mathbb{R}_{+}$, Madlib provides metric differential privacy.

\begin{theorem}\label{th:madlib} For a Euclidean distance metric $d$, Algorithm~\ref{alg:madlib} is $\varepsilon d$-differentially private. Proof in \cite{madlib2020}.
\end{theorem}

Next we develop our algorithm that satisfies metric-DP, giving formal proof of its privacy guarantees.

	\section{Mechanism}

Our proposed mechanism will employ metric differential privacy to sanitize words via their embedding vectors. The challenge with
protecting the privacy is that, for any given input word, the output of a DP algorithm as defined in \cite{dwork2006calibrating} can be any word in the vocabulary, i.e. the
outputs of a mechanism for any pair of inputs words are relatively similar. Metric-DP provides a generalization of DP that allows
adjusting the privacy of an input by leveraging a given distance metric, thus being a suitable framework for improved utility in the text scenario.

However, algorithms like Madlib rely on having access to word embedding vectors and an approximate nearest neighbor index to map noisy vectors to words. The space cost of these can range from hundreds of MB to several
GBs. Thus, using such representations to sanitize words on a user's device would be impractical. In this context, recent
research \cite{tissier2019near, shen2019learning} has focused on converting pre-trained real valued embedding vectors into binary representations. 
They explicitly aim at keeping the semantic meaning while transforming representations. These works show experimental results on
machine learning tasks with the binarization of word embeddings leading to a loss of approximately 2\% in accuracy, with the
upside of reducing the embedding's size by 97\%. Thus, in this work we propose to use binary embeddings of words, obtained from
transforming publicly available continuous representations, such as GloVe \cite{pennington2014glove} or FastText \cite{fasttext2017}.
In addition, specialized nearest neighbor indexes for binary vectors exist \cite{norouzi2012fast}, that dramatically reduce the space and
time cost of nearest neighbor retrieval.

Randomized Response (RR) \cite{warner1965randomized} is a mechanism that provably
\cite{wang2016rr} obtains better utility than the classical Laplace mechanism for binary data collection. RR is a method that dates back to 1965, its original
purpose being to motivate survey respondents to answer questions truthfully, without the risk of exposing any private information. 
For sensitive yes/no questions, survey participants would use a spinner, similar to flipping a coin, and based on its outcome, would either respond truthfully if, for example, the coin came up heads, or respond yes otherwise.
This mechanism provides individuals plausible deniability, while allowing researchers to de-bias the results and obtain the aggregate
metrics they need.
In our case, RR flips a
given input bit of an embedding vector with probability inversely proportional to the privacy parameter $\varepsilon$. We describe a general version
\cite{wang2016rr} of RR on Algorithm~\ref{alg:rr}.

\begin{algorithm}[H]
	\caption{- \textbf{RR}: Randomized Response}
	\label{alg:rr}	
	\begin{algorithmic}[1]
		\INPUT Bit $b \in \{0, 1\}$ and privacy parameter $\varepsilon$.\\
		\OUTPUT Privatized bit $\hat{b}$.
		\STATE Set $\hat{b} = b$ with probability $\frac{e^{\varepsilon}}{1+e^{\varepsilon}}$, otherwise $\hat{b}=  1-b$
		\STATE Return $\hat{b}$
	\end{algorithmic}
\end{algorithm}

RR satisfies metric differential privacy with respect to the privacy parameter $\varepsilon$ and Hamming distance. Since this is
the first time RR is applied to metric-DP, we include a proof of its privacy guarantees.

\begin{theorem}\label{th:rr} For a Hamming metric $d$, Algorithm~\ref{alg:rr} is $\varepsilon d$-differentially private.
	\begin{proof}
		For $RR$ to satisfy metric differential privacy with Hamming distance metric $d$, we have to show that, for two bits $b$ and $b'$ and response bit $y$ we get:
		\begin{gather}\label{eq:proof_rr}
		\frac{\Pr[RR(b)=y]}{\Pr[RR(b')=y]} \leq e^{\varepsilon d(b, b')}
		\end{gather}
		
		For $b = b'$ we have $\Pr[RR(b)=y] = \Pr[RR(b')=y]$ and also $d(b, b') = 0$, therefore Equation~\ref{eq:proof_rr} is satisfied.
		
		For $b \neq b'$, from the definition of RR, we have:
		\begin{gather}
		\nonumber \frac{\Pr[RR(b)=y]}{\Pr[RR(b')=y]} \leq \frac{ \frac{e^{\varepsilon}}{1+e^{\varepsilon}} }{ \frac{1}{1+e^{\varepsilon}} } = e^{\varepsilon}
		\end{gather}
		
		Since for $b \neq b'$ we also get $d(b, b') = 1$, this means we also have Equation~\ref{eq:proof_rr} satisfied for this case.
		
	\end{proof}
\end{theorem}

We now describe our mechanism, denoted as Binary embeddings over Randomized Response (BRR), described in Algorithm~\ref{alg:brr}. BRR uses binary embedding vectors to represent
words and applies RR to make each binary vector differentially private. 

\begin{algorithm}[H]
	\caption{- \textbf{BRR}: Mechanism for Text as Binary Embeddings over Randomized Response}
	\label{alg:brr}
	\begin{algorithmic}[1]
		\INPUT Finite domain $\mathcal{W}$, input word $w \in \mathcal{W}$ and privacy parameter $\varepsilon$.\\
		\OUTPUT Privatized word $\hat{w}$.
		\STATE Compute \textbf{binary} embedding vector $\phi_w = \phi(w)$
		\STATE Perturb word embedding vector using \textbf{Randomized Response} to obtain $\hat{\phi}_w = RR(\phi_w, \varepsilon)$
		\STATE Obtain perturbed word: \\ $\hat{w} = argmin_{y \in \mathcal{W}} \norm{\phi(y) - \hat{\phi}_w}$
		\STATE Return $\hat{w}$
	\end{algorithmic}
\end{algorithm}

With the algorithm described, we state the privacy guarantees of BRR.

\begin{theorem}\label{th:brr} For a Hamming metric $d$, Algorithm~\ref{alg:brr} is $\varepsilon d$-differentially private.
	\begin{proof}
		For embeddings that are independent of the data, the nearest neighbor search is just a post-processing step, thus without
		privacy loss. Therefore we only have to analyze the privacy of releasing the perturbed embedding vector.
		
        Consider $d$ as the Hamming distance and $M$ as the mechanism inside BRR that performs the embedding perturbation, such that
		each word $w \in \mathcal{W}$ has a binary embedding representation $\phi_{w} : \{0,1\}^n$. Then for a given output vector $y :
		\{0,1\}^n$, with $i$'th bit as $y_i$ and any pair of inputs $w, w' \in \mathcal{W}$, with $i$'th bits as $w_i$ and $w'_i$, from
		using RR on a single bit position $i$ we have that:
		\begin{gather}
			\nonumber \frac{\Pr[M(w_i)=y_i]}{\Pr[M(w'_i)=y_i]} \leq e^{\varepsilon d(w_i, w_i')} 
		\end{gather}
		
		For $n$ bits, multiplying probabilities for each of them, we then have:
		\begin{align*}
			\frac{\Pr[M(w)=y]}{\Pr[M(w')=y]} &= \\ \prod\limits_{i=1}^{n} \frac{\Pr[M(w_i)=y_i]}{\Pr[M(w'_i)=y_i]}  &\leq e^{\varepsilon \sum\limits_{i = 1}^{n} d(w_i, w_i')}  = e^{\varepsilon d(w, w')}
		\end{align*}
		where the last step comes from the definition of the Hamming distance.
		
	\end{proof}
\end{theorem}

BRR is similar to the Madlib Mechanism, differing in the use of binary embeddings instead of real-valued, and Randomized Response instead
of the Laplace mechanism. Since RR is proven to be better than Laplace mechanism for binary data \cite{wang2016rr}, if our semantic loss for transforming embeddings
into binary is smaller than the gains of using RR instead of Laplace, then BRR is a promising approach compared to Madlib.

More importantly, our mechanism is suitable to privatize data on-device due to the use of binary embeddings and specialized nearest
neighbor index. On one side,
we have the memory/storage size reduction of using binary embedding vectors, and additionally we gain computational efficiency in the
perturbation with RR, implemented as sampling from a binomial distribution together with a XOR operation, which can be done
efficiently at the hardware level. With these
optimizations, the user would only need to share already privatized data, keeping
ownership of the sensitive data. This is a highly desirable feature, as it allows the use of valuable user data for NLP tasks,
but sensitive information never leaves the user's device.

	\section{Comparing Metric-DP Mechanisms}\label{sec:comparingDPmec}

One issue with a fair evaluation of BRR against Madlib is that they use different distance functions. To solve this we propose
fixing a \emph{privacy ratio} that allows us to obtain similar privacy guarantees, even when two mechanisms use different distance
metrics. Due to space limitations we provide a brief description here and refer the interested reader to Appendix \ref{app:comparingDPmec} for a detailed motivation of the method.

To compare mechanisms with different distance metrics, we consider an estimate of \emph{privacy loss bound} $\varepsilon \cdot \mathcal{P}_d$,
where $\mathcal{P}_d$ is defined as an aggregate distance measurement based on the distances between all possible pairs of words.
In this work, we use either $\mathcal{P}_d^{avg}$ where we average the distances of all pairs, or $\mathcal{P}_d^{max}$ where we use the maximum
distance between any two words, for each mechanism. To fairly compare the privacy of two mechanisms, we equalize their bounds via a privacy
ratio.

\begin{definition}[Method to Fix Privacy] Given two randomized mechanisms $\mathcal{M}_A$ and $\mathcal{M}_B$ both taking inputs from $\mathcal{X}$ to output space $\mathcal{Y}$, satisfying respectively $\varepsilon_A d_A$-DP and $\varepsilon_B d_B$-DP, we denote the privacy ratio as $\mathcal{R}_{d_A, d_B} = \mathcal{P}_{d_A}/\mathcal{P}_{d_B}$. In order to ensure a similar privacy loss on both mechanisms, for any given $\varepsilon_A$ defined for $\mathcal{M}_A$ we need to set:
	\begin{gather}
		\varepsilon_B = \mathcal{R}_{d_A, d_B} \cdot \varepsilon_A
	\end{gather}
\end{definition}

In practice, to compare two mechanisms $\mathcal{M}_A$ and $\mathcal{M}_B$, first we calculate the aggregate distances, e.g. max or average distances between all words, $\mathcal{P}_{d_A}$ and $\mathcal{P}_{d_B}$. Then we obtain the privacy ratio $\mathcal{R}_{d_A, d_B} = \mathcal{P}_{d_A}/\mathcal{P}_{d_B}$. Then for experiments we can simply choose for $\mathcal{M}_A$ any value of $\varepsilon_A$ and then for $\mathcal{M}_B$ we set $\varepsilon_B = \mathcal{R}_{d_A, d_B} \cdot \varepsilon_A$. This implies that $\varepsilon_B \cdot \mathcal{P}_{d_B} = \varepsilon_A \cdot \mathcal{P}_{d_A}$, fixing the estimate of privacy loss for the two mechanisms considered.

By calculating the privacy ratio $\mathcal{R}_{d_A, d_B}$ we can get values $\varepsilon_A$ and $\varepsilon_B$ to obtain similar privacy guarantees of two mechanisms $\mathcal{M}_A$ and $\mathcal{M}_B$. We note that previous work \cite{poincareMDP2019} compared the privacy of two mechanisms looking at the probability that a mechanism returns the exact same word used as input. However, such notion does not always give a fair comparison between mechanisms, as that can vary considerably on embedding spaces. In contrast, our method is theoretically robust, with well defined bounds.
	
	\section{Experiments}

In this section we compare BRR to previous work in terms of privacy, utility and efficiency. Since in DP we usually have a
privacy-utility trade-off when applying mechanisms, we will fix the privacy of our mechanisms in order to compare their utility, using the methodology described in Section \ref{sec:comparingDPmec}.
Our experiments will use the IMDB dataset \cite{imdbData}, with training data being 50\% of the original training dataset, validation data
being the other 50\% of the original training data, and testing data being 50\% of the original testing dataset. The experiments were
performed on an AWS EC2 p3.2xlarge instance. More specific 
setup is described on the next paragraphs.

\textbf{Privacy}: In our experiments we vary $\varepsilon$ for Madlib and set the $\varepsilon$ for BRR using the privacy ratio. To have a more conservative approach, we use an average distance aggregation $\mathcal{P}_d^{avg}$ to fix privacy loss of BRR and Madlib. We note though 
that using $\mathcal{P}_d^{max}$ would give significantly better results for BRR in comparison to Madlib.

\begin{figure}[h]
	\begin{subfigure}{\columnwidth}
		\centering
		\includegraphics[scale=0.5]{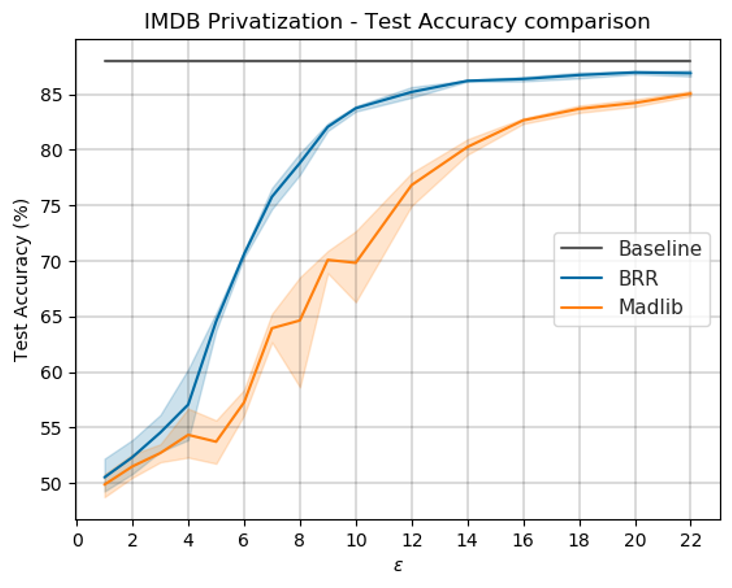}
		\caption{Utility Evaluation}
		\label{fig:imdb_utility}
	\end{subfigure}%
    \\
	\begin{subfigure}{\columnwidth}
		\centering
		\includegraphics[scale=0.5]{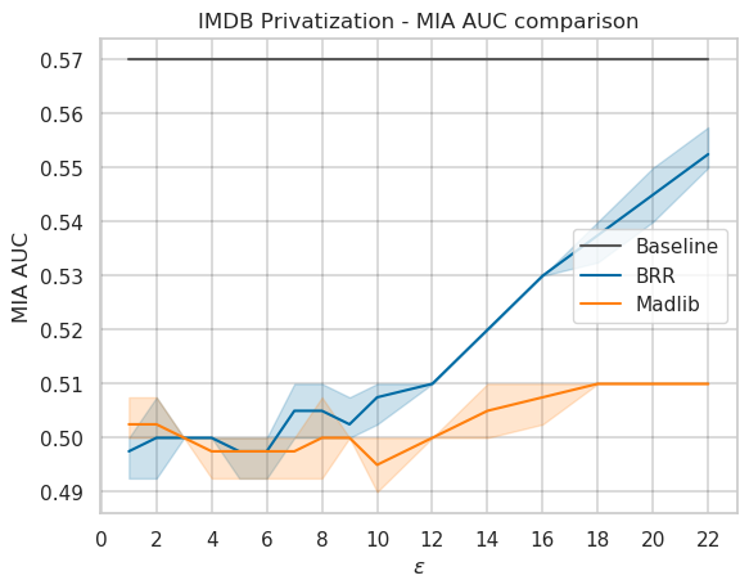}
		\caption{MIA Evaluation}
		\label{fig:imdb_mia}
	\end{subfigure}%
	\caption{Comparison of mechanisms with 95\% confidence interval over 10 independent trials for various $\varepsilon$. Baseline is
		built with models trained on original sensitive data. $\varepsilon$ is first set for Madlib and transformed for BRR using the method from Section \ref{sec:comparingDPmec}.}
	\label{fig:imdb}
\end{figure}

\textbf{Utility}: To evaluate the utility of the metric-DP mechanisms, we build ML models for sentiment analysis on training data privatized by each
mechanism and compare the accuracy of the trained models on a separate testing dataset. For Madlib we use Euclidean distance on a
fixed embedding space from GloVe \cite{pennington2014glove} with 300 dimensions. BRR starts from the same embedding, then
transforms it into a binary representation with 256 dimensions as described in \cite{tissier2019near}. The sentiment classification models follow the FastText classifier \cite{Joulin_2017}. The accuracy on the test dataset is shown in Figure~\ref{fig:imdb_utility}.

As seen in Figure \ref{fig:imdb_utility}, in terms of utility, we obtain similar results for Madlib's $\varepsilon$ under 5, and improved results for
BRR for larger $\varepsilon$, where each setting is averaged over 5 independent trials, with the shaded are showing the 95\%
confidence interval.

\textbf{Defense against attacks}: We also include the results of a Membership Inference Attack (MIA) \cite{shokri2017membership}, which tries to infer the presence of observations used to train a given model based only on black-box access, that is, the attacker is only able to query the deployed model and does not have access to specific weights. Lower score of the attack is better, representing more privacy preservation. 

In summary, we have a \textit{target} model that we can only query, e.g. through a public API, which we plan to attack to determine membership of particular users in its training set. To attack the target model we train a \textit{shadow} model, based on disjoint data we have available. This model tries to emulate the behavior of the target model. For example, we could try attacking a next-word prediction model trained on private data, by training a similar model using data from a user's public Twitter feed. Our dataset will then have two labels. The first called \textit{member} we use to train the shadow model. The second, called \textit{non-member} we set aside to train an \textit{attack} model that will tell us if an example was part of the training data of the target model.

For this step, we use 50\% of the original IMDB testing data to train a shadow model, another 50\% to validate it. Models created for utility evaluation
are the target for the attack. The attack model is an MLP with two layers having 64 hidden nodes each, with ReLU activations. Results are on Figure~\ref{fig:imdb_mia}. We can see that for the various privacy levels represented by $\varepsilon$ tested, we obtain practical privacy protection,
represented by the drop in AUC of the attack model. More specifically, we see that as we increase $\varepsilon$, which decreases the formal privacy guarantees of DP, we also obtain also less \textit{empirical} privacy, represented by larger AUC. In this case, for very large $\varepsilon > 12$ we see a bigger impact on MIA for BRR. However, even though Madlib has smaller impact seen on MIA, that is not a \textit{formal} privacy guarantee. Therefore, such large values of $\varepsilon$ are not recommended for both mechanisms on the dataset analyzed. Finally, we note that for $\varepsilon \leq 12$, where both mechanisms have $\text{AUC}$ of approximately $0.50$ for MIA, we see BRR with similar or better utility than Madlib.

\textbf{Efficiency}: To evaluate how efficient the mechanisms are on different aspects, we consider the size of embedding vectors and
index for approximate nearest neighbors, the wall time of privatization per word and total wall time per mechanism. For nearest
neighbors search, we use FAISS~\cite{faissJDH17}, a library that deals with both real and binary vectors.

For our relatively small vocabulary, the nearest neighbors index built for BRR achieved a compression rate of 97.9\% (4MB vs. 200MB), while the vocabulary
file, along with embeddings, was also 98.5\% smaller (6MB vs. 300MB) compared to the ones used by Madlib.

\begin{figure}[h]
	\centering
	\includegraphics[width=.45\textwidth]{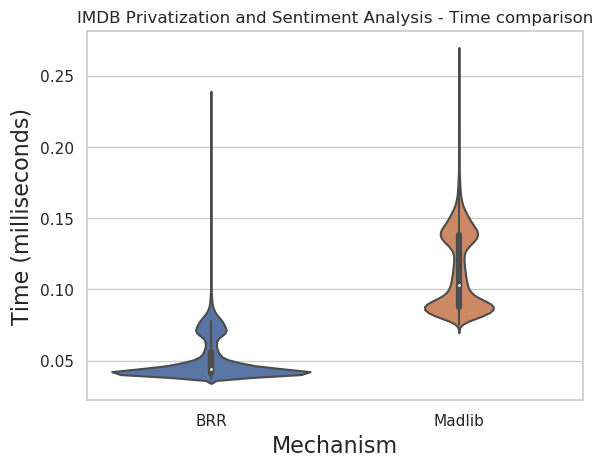}
	\caption{Wall-time comparison between BRR and Madlib on IMDB dataset.}
	\label{fig:time_comp}
\end{figure}


For a fair time comparison, we looked for optimal ways of improving the computation of nearest neighbors during
privatization of both BRR and Madlib. This step is the most time consuming of both methods and Madlib uses real-valued data with
the Euclidean distance, while BRR uses binary data and the Hamming distance.

In our experiments, BRR was on average 68\% faster to privatize a word
compared to Madlib, as we can see on Figure~\ref{fig:time_comp}. 
Using binary embedding vectors significantly improved the running time of our solution compared to using real-valued vectors, due to
the possibility of using more efficient algorithms and hardware-level optimizations through binary operations. 
We used the FAISS library\footnote{\url{https://github.com/facebookresearch/faiss}}
that implements algorithms \cite{norouzi2012fast} tailored for binary data, for both exact and
approximate nearest neighbors search. We note that FAISS has additional features in the library that we did not apply but could
be further explored, such as product quantization, which is a technique for lossy compression of
high-dimensional vectors, and PCA. Similarly, for Madlib we use approximate nearest neighbors, more specifically
the Annoy library\footnote{\url{https://github.com/spotify/annoy}}. This speeds the retrieval at the cost of losing the guarantee 
to find the exact nearest neighbor.

	\section{Conclusion}

We presented a mechanism for efficient text privatization on-device with formal differential privacy guarantees.
We demonstrated that our new mechanism enables performing privatization on-device through the use of binary embeddings, providing zero-trust
privacy for customers, while maintaining better or competitive utility than the state-of-the-art.
As future work, we plan to implement improvements in the nearest neighbors search step, exploring smaller dimensionality
of embedding vectors and including privatization of word context vectors for improved utility. We are currently exploring other DP mechanisms
like using the exponential mechanism\cite{mt2007expmech} that allows for the use of any metric function in the privatization process. Finally, a related line of work is how to achieve on-device privacy for tasks like Automatic Speech Recognition. 
	
	\bibliographystyle{plainnat}
	\bibliography{references.bib}
	
	\clearpage
	
	\appendix
	
	\section{Motivation for Privacy Ratio}\label{app:comparingDPmec}

In the standard definition of differential privacy~\cite{dwork2006calibrating}, the privacy guarantees provided by different mechanisms are compared by looking only at the $\varepsilon$ value: mechanisms with same $\varepsilon$ provide the same privacy guarantees. However, referring to Equation~\ref{eq:mDP} we see that for metric-DP the privacy bound is directly impacted not only by $\varepsilon$ but also by the distance metric. As a result, comparing mechanisms with different metrics is a non-trivial task. In this context we propose calculating, for each mechanism, a privacy loss bound for all $x, x' \in \mathcal{X}$.

To compare privacy among mechanisms first we look at the privacy loss, which is a general function defined for randomized mechanisms, not specific to any differential privacy definition.

\begin{definition} Let $\mathcal{M} : \mathcal{X} \rightarrow \mathcal{Y}$ be a randomized mechanism with density function $p_{\mathcal{M}(x)}(y)$, then for any $x, x' \in \mathcal{X}$ and all outputs $y \in \mathcal{Y}$ the privacy loss function is defined as:
	\begin{gather}
		\mathcal{L}_{\mathcal{M}, x, x'}(y) = \ln\bigg( \frac{p_{\mathcal{M}(x)}(y)}{p_{\mathcal{M}(x')}(y)} \bigg)
	\end{gather}
	
\end{definition}

Every differentially private mechanism will have by definition an upper bound for $\mathcal{L}_{\mathcal{M}, x, x'}(y)$, and more specifically, metric-DP, as we show next.

\begin{corollary} Given a randomized mechanism $\mathcal{M}: \mathcal{X} \rightarrow \mathcal{Y}$ that is $\varepsilon d$-differentially private for a given distance function $d: \mathcal{X} \times \mathcal{X} \rightarrow \mathbb{R}_{+}$, we have that for any $x, x' \in \mathcal{X}$ and all outputs $y \in \mathcal{Y}$:
	\begin{gather}\label{eq:plbound}
		\mathcal{L}_{\mathcal{M}, x, x'}(y) < \varepsilon \cdot d(x, x')
	\end{gather}
\end{corollary}

From Equation~\ref{eq:plbound} above we can see that the privacy loss bound of any pair $x,x'$ depends on their distance and $\varepsilon$. Therefore, for a given metric $d$ we propose calculating $d(x,x')$ for every possible pair $x,x' \in \mathcal{X}$ in order to obtain the overall bound on the privacy loss. With these distances, we can define a privacy measurement $\mathcal{P}_d$ for the maximum or average, as we now formalize.

\begin{definition} For a given distance function $d: \mathcal{X} \times \mathcal{X} \rightarrow \mathbb{R}_{+}$ and finite space $\mathcal{X}$, we define the privacy measurement $\mathcal{P}_{d}^{max}$ and $\mathcal{P}_d^{avg}$ as:
	\begin{gather}\label{eq:pd_max}
		\mathcal{P}_d^{max} = \max_{\forall x, x' \in \mathcal{X}} d(x, x')
	\end{gather}
	\begin{gather}\label{eq:pd_avg}
		\mathcal{P}_d^{avg} = \sum_{\forall x, x' \in \mathcal{X}} d(x, x') / | \mathcal{X} |^2
	\end{gather}
\end{definition}

For example, in the context of words as input, $\mathcal{P}_d$ can be calculated using a given distance metric on the embedding space for every word in the vocabulary. Comparing the two privacy measurements above, we note that $\varepsilon\cdot \mathcal{P}_d^{max}$ has the advantage of giving an overall bound on the privacy loss. This can be derived from Equations~\ref{eq:plbound} and \ref{eq:pd_max}, such that for any $x, x' \in \mathcal{X}$ and all outputs $y \in \mathcal{Y}$, the privacy loss is bounded by:
\begin{gather}\label{eq:pl_exp_bound}
	\mathcal{L}_{\mathcal{M}, x, x'}(y) < \varepsilon \cdot d(x, x') < \varepsilon \cdot \max_{\forall x, x' \in \mathcal{X}} d(x, x') = \varepsilon \cdot \mathcal{P}_d^{max}
\end{gather}

On the other hand, $\mathcal{P}_d^{avg}$ gives average distance bounds over a given space $\mathcal{X}$, which is a more conservative approach and may be of interest to avoid spaces with outliers that individually would largely affect the privacy loss upper bound.

Given the above, $\varepsilon \cdot \mathcal{P}_d^{avg}$ or $\varepsilon \cdot \mathcal{P}_d^{max}$ can be used as estimates of privacy loss bounds, which allows us to fairly compare mechanisms using different distance metrics, as defined in Section \ref{sec:comparingDPmec} by equalizing their bounds through a privacy ratio.

\end{document}